\documentclass[twocolumn,a4,prb]{revtex4} 
\usepackage{graphicx}%
\usepackage{color}
\usepackage{dcolumn} 
\usepackage{bm} 
\makeatletter \def\btt#1{\texttt{\@backslashchar#1}}%
\DeclareRobustCommand\bblash{\btt{\@backslashchar}}%
\makeatother  
\begin{document}  
\preprint{SOLITONCMR.TEX}  
\title{$^{139}$La NMR evidence for phase solitons in the ground state of overdoped manganites}
\author{D. Koumoulis$^1$ , N. Panopoulos$^1$, A. Reyes$^2$, M. Fardis$^1$, M. Pissas$^1$, A. Douvalis$^3$, T. Bakas$^3$, D. Argyriou$^4$ \& G. Papavassiliou$^{1*}$} 
\affiliation{$^1$Institute of Materials Science, NCSR,  Demokritos, 153 10 Aghia Paraskevi, Athens, Greece}
\affiliation{$^2$National High Magnetic Field Laboratory, Tallahassee, Florida 32310, USA}
\affiliation{$^3$Physics Department, University of Ioannina, P.O. Box 1186, GR-451 10, Ioannina, Greece}
\affiliation{$^4$Helmholtz-Zentrum Berlin fuer Materialien und Energie (HZB), Glienicker Strasse 100, D-14109}

\date{\today }   
\maketitle  
{\bf 
Hole doped transition metal oxides are famous due to their extraordinary charge transport properties, 
such as high temperature superconductivity (cuprates) and colossal magnetoresistance (manganites) \cite{Dagotto05}. Astonishing, the mother system of these compounds is a Mott insulator \cite{Emery99}, whereas important role in the establishment of the metallic or superconducting state is played by the way that holes are self-organized with doping \cite{Emery99,Tranquada95,Tranquada94}. Experiments have shown that by adding holes the insulating phase breaks into antiferromagnetic (AFM) regions, which are separated by hole rich clumps (stripes) with a rapid change of the phase of the background spins and orbitals \cite{Emery99,Zaanen89}. However, recent experiments in overdoped manganites of the La$_{1-x}$Ca$_x$MnO$_3$ (LCMO) family have shown that instead of charge stripes \cite{Chen96,Chen97,Mori98}, charge in these systems is organized in a uniform charge density wave (CDW) \cite{Loudon05,Nucara08,Milward05}. Besides, recent theoretical works predicted that the ground state is inhomogeneously modulated by orbital and charge solitons \cite{Brey04,Brey05}, i.e. narrow regions carrying charge $\pm e/2$, where the orbital arrangement varies very rapidly. So far, this has been only a theoretical prediction. Here, by using $^{139}$La Nuclear Magnetic Resonance (NMR) we provide direct evidence that the ground state of overdoped LCMO is indeed solitonic. By lowering temperature the narrow NMR spectra observed in the AFM phase are shown to wipe out, while for $T<30$K a very broad spectrum reappears, characteristic of an incommensurate (IC) charge and spin modulation. Remarkably, by further decreasing temperature, a relatively narrow feature emerges from the broad IC NMR signal, manifesting the appearance of a solitonic modulation as $T\rightarrow 0$.
}

The presence of an IC spin-density modulation and the formation of charge stripes in AFM transition metal oxides was first observed in superconducting La$_{2-x}$Sr$_x$CuO$_4$ and their insulating nickelate counterparts \cite{Cheong91,Hayden92}. Inelastic neutron scattering experiments have shown that holes in these materials tend to localize into periodically arranged antiphase domain walls, separating AFM regions, which propagate diagonally through the CuO$_2$ (respectively NiO$_2$) layers \cite{Emery99}. In addition, a link between stripe ordering and the wipeout effect (disappearance) of the NMR/NQR signal was observed \cite{Hunt99,Curro00}, which indicates that most probably the stripe phase in these systems is a slowly fluctuating, strongly correlated fluid over an extended temperature range. 

In case of the LCMO family for $x\geq 0.5$, electron diffraction experiments unveiled an IC charge modulation with wave vector $q\parallel a^{\star }$ \cite{Chen96,Chen97,Mori98}, which is associated with orbital and AFM spin ordering. This charge modulation has been considered as signature of charge stripes arising from the ordered arrangement of alternating Mn$^{+3}$ and Mn$^{+4}$ ions. However, a number of recent experiments put into question this kind of Mn$^{+3}$ and Mn$^{+4}$ charge alternation \cite{Subias97,Martin04}; instead they suggest the formation of a charge modulation wave with a uniform periodicity for all $x \geq 0.5$ \cite{Loudon05}. A collective sliding of the charge system in a moderate electric field was also reported at half doping, $x =0.5$ \cite{Cox07}. These experimental results have been explained in the framework of a CDW model \cite{Milward05,Mathur01} with nonzero and possibly high itineracy. Recently, it has been also proposed that the nanoscale spin and orbital texture observed in overdoped manganites is possibly explained in the framework of an IC solitonic ground state \cite{Brey04,Brey05}, that is produced by orbital solitons carrying charge equal to $\pm e/2$ \cite{Brey05}. However, until now there is no experimental evidence about the existence of orbital or spin solitons; only a uniform IC modulation has been reported to exist in overdoped LCMO for $T\geq 90$K \cite{Loudon05}. Here, by using $^{139}$La NMR in magnetic field $4.7$ Tesla, we provide clear evidence about the formation of a uniform IC spin and charge modulation in overdoped LCMO manganites, which becomes visible for $T<30$K. Most important, at very low temperatures this modulation is shown to split into regularly arranged commensurate and narrow IC (soliton) regions, pointing at an IC soliton-modulated spin and charge ground state. 

\begin{figure} [htb]
\centering
\includegraphics[angle=0,width=8cm]{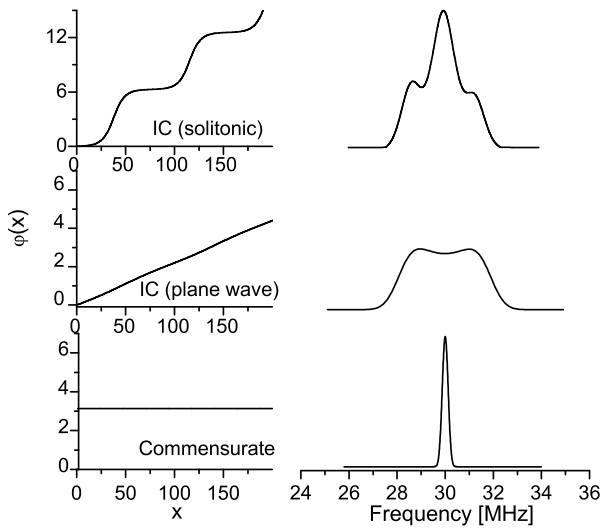}
\caption{The spatial variation of the phase $\varphi (x)$ of a SDW and the corresponding NMR lineshapes for three different cases: (i) A spin modulation with constant phase $\varphi (x)$ (lower panel), corresponding to a commensurate spin configuration with the crystal-lattice periodicity. (ii) An IC plane wave phase modulation (middle panel), and (iii) an IC modulation with phase solitons (upper panel).}   
\label{fig1} 
\end{figure}

Until now, the most detailed information about charge, spin and orbital modulated phases in doped AFM transition metal oxide compounds has been acquired by neutron and electron scattering techniques \cite{Tranquada95,Chen97}. Diffraction of a neutron or electron beam by long-period spin and charge density modulations yields sattelite Bragg peaks, which provide information about the spatial period and orientation of the corresponding modulation, whereas the intensity provides a measure of the modulation amplitude. A complementary and equally powerful technique in the study of modulated structures is NMR. In the past, NMR has been succesfully applied in the study of structural IC phases \cite{Blinc81}, as well as CDW \cite{Berthier76} and SDW systems \cite{Takahashi86}. The ability of NMR to provide information about the appearance and evolution of an IC spin modulation is clearly demonstrated in Figure \ref{fig1}.  

Assuming an $1-D$ spin density wave (SDW), the modulation wave varies in space according to formula $\rho =A\cos(\varphi (x))$ \cite{Milward05,Mathur01}. In magnetic systems the NMR frequency is proportional to the local hyperfine field $B_{hf}=(1/\gamma \hbar )C{<S>}$, (C is the hyperfine coupling constant and ${<S>}$ the average electron spin probed by the resonating nuclei \cite{Papavassiliou01}); hence, the NMR frequency will be accordingly modulated, $\nu =\nu _0+\nu _1\cos(\varphi (x))$. This gives rise to a characteristic NMR frequency distribution depicted by formula, $f(\nu )\propto  \frac{1}{\nu_1 \mid \sin(\varphi)\frac{d\varphi}{dx}\mid }$ \cite{Blinc81}. In case of an IC modulation, the phase of the modulation wave $\varphi (x)$ varies in space according to the Sine-Gordon (SG) equation, $\frac{d^2\varphi}{dx^2}=w\sin(m \varphi(x)) $ \cite{Mathur01,Rice76,Horovitz78}, where $m$ is a commensurability factor defining the crystal symmetry \cite{Mathur01,Rice76}, and $w$ a constant depending on the electronic properties of the system and the amplitude of the modulation wave \cite{Mathur01,Horovitz78}. Figure \ref{fig1} presents the spatial variation of  $\varphi (x)$ by solving the SG equation for two cases: (i) A nearly plane wave IC modulation with $w<<1$ (left middle panel), and (ii) an IC solitonic phase modulation with $w\leq 1$ (left upper panel). The commensurability factor was set equal to $m=1$, in accordance with theoretical studies for $1-D$ CDWs \cite{Horovitz78}. This gives a phase shift accross solitons equal to $\Delta \varphi =2\pi $. Other works predict $\Delta \varphi =\pm \pi/2$ and a fractional charge $\pm e/2$ \cite{Brey05}, while in systems with coexisting charge and spin order a change in phase equal to $\pm \pi $ has been predicted, even in the presence of higher order commensurability (e.g. $m=3$ or $4$) \cite{Horovitz82}. For reasons of comparison a commensurate modulation with constant phase $\varphi (x)$ is also presented in the left lower panel of Figure \ref{fig1}, which corresponds to spin configuration with the crystal-lattice periodicity. Simulated NMR spectra for the three cases (with initial phase $\varphi_0=\pi /2$ for the modulation wave) are presented in the right panels of Figure \ref{fig1}. Evidently, in case that the system undergoes a series of Commensurate$\rightarrow$IC (plane wave)$\rightarrow$IC(solitonic) phase transitions, the NMR signal is expected to transform from a narrow symmetric lineshape in the commensurate phase, to a broad frequency distribution in the IC plane wave limit, and finally to the composite signal of the right upper panel. 

\begin{figure} [htb]
\centering
\includegraphics[angle=0,width=8cm]{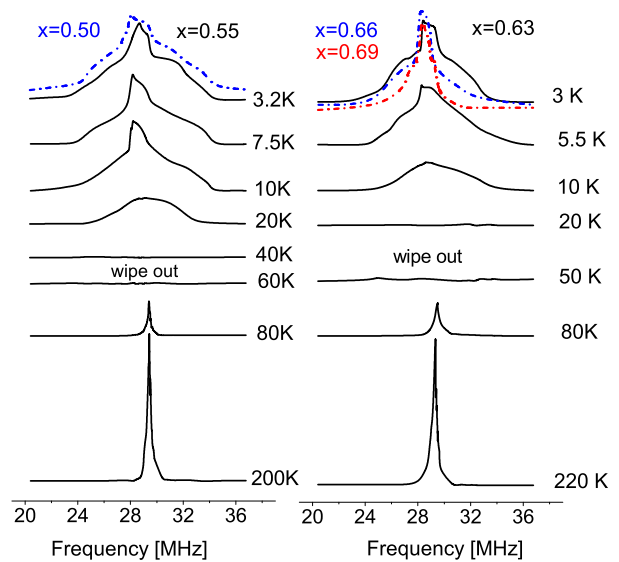}
\caption{$^{139}$La NMR lineshapes of the AFM signal as a function of temperature for $x=0.55$ and $0.63$. The blue line in the left panel is the spectrum for $x=0.50$ at $4$K, and the blue (red) lines in the left panel are the spectra for $x=0.66$ ($0.69$) at $3$K.}
\label{fig2} 
\end{figure}

Figure \ref{fig2} shows $^{139}$La NMR lineshapes for LCMO with $x=0.55$ and $0.63$ in the temperature range $3$K to $200$K. A striking similarity is observed between the simulations of Figure \ref{fig1} and the temperature evolution of the spectra in Figure \ref{fig2}. For $80$K$\leq T\leq 200$K the NMR spectra consist of a narrow symmetric line at frequency $\approx 29$ MHz, which remains almost unshifted by entering the AFM phase ($T_N\approx 150$K). This is reminiscence of an AFM spin modulation that is commensurate with the underlying crystal lattice. Remarkably, for $T<T_N$ the signal intensity decreases rapidly by cooling and disappears at $\approx 60$K. This impressive wipe-out effect is similar to the one observed in high $T_c$ cuprates and nickelates \cite{Hunt99,Curro00}, and will be discussed in more detail below. Unexpectedly, a very broad spectrum reappears for $T\leq 30$K , which is characteristic of an IC modulation. By further lowering temperature a relatively narrow peak emerges from the broad IC NMR signal, which closely resembles the spectrum in the upper right panel of Figure \ref{fig1}. This is strongly suggesting that by lowering temperature, the CDW/SDW breaks into alternating commensurate and discommensurate regions giving rise to the particular soliton-modulated NMR signal. A similar temperature evolution of the NMR spectra was observed in LCMO $x=0.50$, $0.66$, and $0.69$. However, the IC modulation appears to reduce at higher doping values, as indicated by the narrowing of the IC NMR lineshapes with increasing doping in Figure \ref{fig2}.  This is in conformity with electron diffraction experiments, which exhibit a decrease of the modulation wavevector according to relation, $q_s=(1-x)a^{\ast }$  \cite{Loudon05a}. 

\begin{figure} [htb]
\centering
\includegraphics[angle=0,width=8cm]{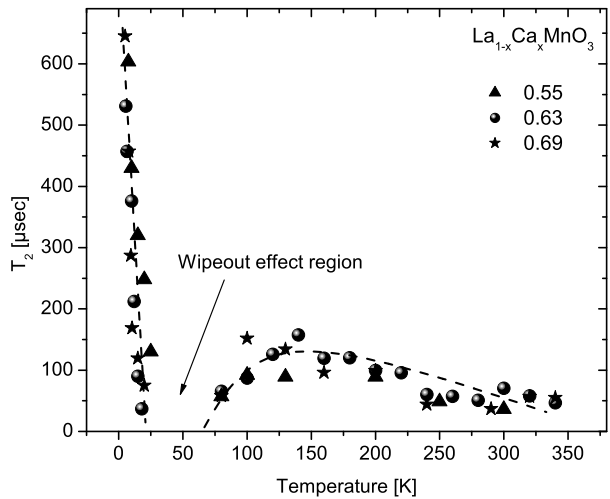}
\caption{$^{139}$La NMR spin-spin relaxation time ($T_2$) as a function of temperature of LCMO $x=0.55, 0.63$ and $0.69$. The temperature range without experimental points ($30-70$K) is the range where the NMR signal disappears, due to extremely short $T_2$ values (wipeout).}
\label{fig3} 
\end{figure}

Another distinguishing feature in Figure \ref{fig2} is the strong wipeout effect of the NMR signal, which takes place in the temperature range $30$K to $60$K. The reason for the complete disappearance of the NMR signal is the extreme shortening of the spin-spin relaxation time $T_2$, as shown in Figure \ref{fig3}. This makes part of the La nuclei to relax so fast that the NMR signal decays before it can be measured. In general, the $T_2$ shortening by cooling is ascribed to slowing down of CDW/SDW fluctuations in accordance with relevant measurements in other systems \cite{Hunt99,Curro00,Papavassiliou01}. The presence of slow collective fluctuations reconciles with resistivity measurements, where application of a moderate external electric field invokes sliding of the CDW \cite{Cox07}. 

\begin{figure} [htb]
\centering
\includegraphics[angle=0,width=8cm]{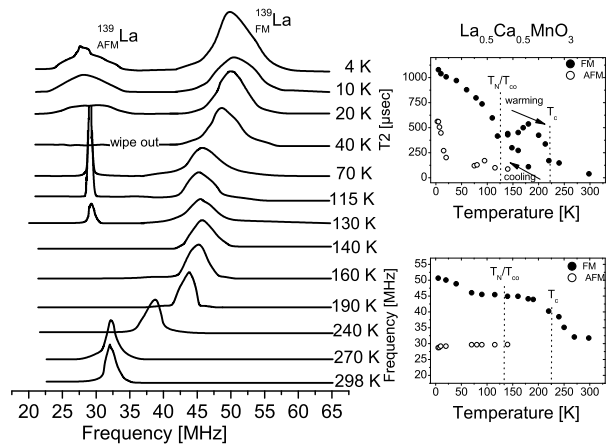}
\caption{$^{139}$La NMR spectra for $x=0.50$ at various temperatures. For $T<70$K the AFM signal component wipes out, while for $T<30$K the broad IC NMR signal marks the appearance of an IC modulated AFM phase. The upper right panel shows the spin-spin relaxation time $T_2$ as a function of temperature for both the FM and AFM signal components. The lower right panel shows the corresponding signal frequencies.} 
\label{fig4} 
\end{figure}

Of considerable interest is the $x=0.5$ sample, which lies at the phase boundary of the $T-x$ magnetic phase diagram, where coexistence of a relatively strong FM phase component with the AFM phase is observed. By lowering temperature this system undergoes a Paramagnetic (PM) to Ferromagnetic (FM) phase transition at $T_c\approx 220$K and a first order FM to AFM phase transition at $T_N\approx 180$K by cooling (respectively $T_N\approx 130$K by heating) \cite{Radaelli95}. Electron \cite{Chen96} and neutron \cite{Radaelli95} diffraction experiments revealed the formation of a commensurate charge modulation for $T<T_N$, which becomes IC at higher temperatures. Figure \ref{fig4} shows the evolution of the $^{139}$La NMR signal from room temperature down to $3$K. At room temperature the PM NMR signal is located at $\approx 32$MHz. By lowering temperature, the NMR signal starts to shift at higher frequencies at $\approx 260$K, demonstrating the transition from the PM to the FM phase (the transition is shifted at higher temperatures in magnetic field $4.7$ Tesla). At temperatures lower than $T_N$ the AFM signal component starts to grow at $\approx 29$ MHz, which exhibits the same behaviour as in all other studied samples; the narrow AFM NMR signal wipes out at $\approx 60$K, whereas a very broad IC signal reappears followed by a solitonic IC spectra at very low temperatures. This is a very important result because it rules out a perfectly commensurate zig-zag charge and spin ordering at low temperatures \cite{Chen96,Radaelli95}. The upper right panel of Figure \ref{fig4} shows $T_2$ vs. $T$. As expected, a strong hysterestic behaviour between $T_c$ and $T_N$ is shown by the $T_2$ of the FM signal component. On the other hand, the $T_2$ of the AFM signal component exhibits the same behaviour as in all other systems presented in Figure \ref{fig3}. We also notice that the high temperature AFM signal is sufficiently narrower than the PM signal, which supports the idea that in this temperature regime the NMR signal is motionally narrowed, due to strong fast fluctuations of the modulation wave. Another important remark is that the onset of the wipeout effect is accompanied with a strong frequency shift and significant broadening of the FM NMR signal. This indicates that the FM phase component does not consist of isolated FM islands embedded into an AFM matrix, but is rather strongly interacting with the AFM phase component. 

In summary,  our $^{139}$La NMR measurements provide evidence that the ground state in overdoped LCMO manganites comprises an IC soliton-modulated charge and spin density wave. At higher temperatures the modulation wave transforms to a uniform IC plane wave, which is subjected to strong slow fluctuations, as implied by the complete wipeout effect of the NMR signal. This is a completely new result, which urges to reconsider our opinion about the low temperature electronic properties of overdoped manganites. Even more, the fundamental mechanism governing the establishment and evolution of the stripe phase appears to be common in overdoped manganites with cuprates and nickelates. In all these systems charge order is established at a higher temperatures than AFM order, while the growth of the IC modulation at low temperatures is accompanied with a strong wipeout of the NMR signal. Further experiments are needed in order to examine whether the solitonic ground state is a generic property of striped AFM transition metal oxide compounds.


\end{document}